\documentclass[aps,pra,twocolumn,showkeys,amsmath,preprintnumbers,amssymb,superscriptaddress]{revtex4-1}
\usepackage{amsmath}
\usepackage{graphicx}
\usepackage[latin1]{inputenc}
\usepackage{array}
\newcolumntype{L}{>{\displaystyle}l}
\newcolumntype{C}{>{\displaystyle}c}
\newcolumntype{R}{>{\displaystyle}r}
\def\>{\rangle}
\def\<{\langle}

\graphicspath{{./fig/}{./pdf/}}

\begin{document}


\title{Resolubility of Image-Potential Resonances}
\date{\today}
\author{Ulrich H\"ofer}
 \affiliation{Fachbereich Physik und Zentrum f{\"u}r
Materialwissenschaften, Philipps-Universit{\"a}t, 35032 Marburg, Germany}
\affiliation{Donostia International Physics Center (DIPC), 20018 San
Sebasti{\'a}n, Spain}
\author{Pedro~M.~Echenique}
\affiliation{Donostia International Physics Center (DIPC), 20018 San
Sebasti{\'a}n, Spain} \affiliation{Departamento de F{\'\i}sica de
Materiales, UPV/EHU and CFM-MPC, Centro Mixto CSIC-UPV/EHU, San
Sebasti\'an}

\date{\today}
\keywords{Image-potential states, surface resonances, open quantum
system}

\begin{abstract}                                                   %
%
A theory of image-potential states is presented for the general case
where these surface electronic states are resonant with a bulk
continuum.
 The theory extends the multiple scattering approach of Echenique and Pendry
into the strong coupling regime while retaining independence from
specific forms of surface and bulk potentials.
 The theory predicts the existence of a well-resolved series of resonances for
arbitrary coupling strengths. Surprisingly, distinct image-potential
resonances are thus expected to exist on almost any metal surface,
even in the limiting case of jellium.
\end{abstract}
\maketitle
%

\subsection*{1.~Introduction}

 At surfaces and interfaces of metals, atomic, molecular or other
discrete electronic levels couple to a continuum of states in the
volume.
 The accurate physical description of this coupling is
central to understanding a wide variety of basic processes in catalysis, nanoscience or molecular
electronics.
 Here, we consider one of the most simple, yet fundamental model systems of this kind,
image-potential states, resonant with a structureless continuum.

Experimental and theoretical studies of the ultrafast dynamics of
electrons in image-potential states have vastly improved our
understanding of electronic excitation and decay processes at
surfaces of metals \cite{Bovens10}.
 Electrons excited to these states experience the Coulombic image
force perpendicular to the metal surface. For crystallographic faces
which exhibit a gap of the projected bulk bands in the vicinity of
the vacuum energy, this gives rise to a Rydberg series of states,
characterized by hydrogen-like wavefunctions in the vacuum and
exponentially decaying Bloch waves in the bulk
\cite{Echeni78jp,Fauster95}.
 The conceptual simplicity of these electronic states and their
well-defined properties have not only allowed to identify and
quantify the important factors that govern their decay
\cite{Echeni04ssr,Fauster07pss}.
 In fact, they have attained the role of a kind of drosophila of electron dynamics.
As such, they serve as a benchmark system for the development of new ultrafast
experimental techniques and as a reference for investigations of more complex
electron transfer processes
\cite{Hofer97sci,Ge98sci,Miller02sci,Boger04prl,Gudde07sci,Schwalb08prl,Zhu09acr,Borca10prl,Eickhoff11prl,Cui14natphys}.

 Recently, several experiments showed the existence of a well-defined
series of similar states in the absence of a projected band gap
\cite{Eickhoff11prl,Winter11prl,Marks11prb2}.
 Such image-potential resonances are depopulated mainly by elastic
electron transfer into the bulk whereas the classical
image-potential states can only decay inelastically.
 Since resonant elastic channels are expected to dominate electron
transfer processes at interfaces in most applications, model studies
of the decay of image-potential resonances are most interesting.
These perspectives are, however, seriously hampered by the lack of a
rigorous theoretical description which goes beyond appropriately
tuned model potentials \cite{Borisov06prb,Tsirkin13prb}.
 The multiple scattering theory of Echenique and Pendry \cite{Echeni78jp} to
free-electron-like metals as well as density functional theory at the level of
the GW approximation predict just one broad resonance
\cite{Papadia90prb,Fratesi03prb}.
 These results, which had been well accepted for many years, are now
contradicted by recent experiments for Al(100) \cite{Winter11prl}.

 In this paper, we extend the theory of Ref.~\cite{Echeni78jp} in
order to allow all scattering channels to interfere. It will be shown that such interference
effects decisively change the solution in the regime of strong coupling leading to the full
resolubility of the Rydberg series beyond the $n=1$ state.

\subsection*{2. Theoretical Model}

 We follow ideas first introduced by Feshbach for problems in nuclear
physics \cite{Feshbach58} and apply an {\em open} quantum system
formalism. In this approach one considers the following
non-hermitian effective Hamiltonian
\begin{equation}
\mathcal{H}_{\rm eff} = \mathcal{H}_0 - i VV^{\dagger}.
\label{Heff}
\end{equation}
$\mathcal{H}_0$ describes the unperturbed states.
 In practice, it is a $N\times N$ matrix with diagonal elements $E_n$ and all
off-diagonal elements equal to zero (no configuration interaction).
 $V$ is a $K \times N$ matrix that describes the coupling to $K$
continuum channels.
 The eigenvalues $\lambda$ of $\mathcal{H}_{\rm eff}$ are complex,
with $E_{\rm res}= {\rm Re}(\lambda)$ denoting the maxima of the
density of states and $\Gamma_{\rm res}= -2 {\rm Im}(\lambda)$ their
widths.

 This type of {\em open} quantum system formalism has successfully been
used to solve problems in various fields of physics \cite{Okolow03,Desout95jp}. Gauyacq and
co-workers have applied it to interpret results they obtained for the interaction of two atomic
helium levels as a function of the distance from an aluminum surface \cite{Makhme94el} and, more
recently, to explain the long excited-state lifetime of a metallic double chain adsorbed on
Cu(111) \cite{Diazte09prl}.

 Before we apply the formalisms to the series of image-potential resonances, it is
instructive to simply consider two levels coupled to one continuum channel,
\begin{equation}
 \mathcal{H}_{\rm eff} =
  \left(\begin{array}{CCC}
    -\frac{E_0}{2}  && \\[-3pt]
      && +\frac{E_0}{2}
   \end{array}\right) - i\alpha
  \left(\begin{array}{CCC}
    1 && \sqrt{f} \\[2pt]
    \sqrt{f}  && f
   \end{array}\right).
\label{Heff2}
\end{equation}
 Without loss of generality the two levels are assumed to have
energies $\mp E_0/2$, the lower one with stronger coupling and the
upper one with a coupling that is smaller by a factor of $f$. The
parameter $\alpha$ is a measure of the overall coupling strength of
both levels to the continuum.

One notices that the coupling matrix $i VV^{\dagger}$ in $\mathcal{H}_{\rm eff}$ (\ref{Heff2})
does not only contain the diagonal elements $i\alpha$ and $i\alpha f$ describing a decay. It also
contains the off-diagonal coupling terms $i\alpha \sqrt{f}$. As shown explicitly in the Appendix,
the resulting interference is mediated by the continuum.
 In the coupled system, not only the two levels are affected by the continuum. Similarly, the
initially structureless continuum is disturbed upon interaction with the two levels and this
disturbance, in turn, acts back on the resonances. Although $\mathcal{H}_{\rm eff}$ does not
describe the continuum itself, it includes this interference effect in leading order.

The complex eigenvalues of (\ref{Heff2}) are
\begin{equation}
 \lambda_{1,2} =
 \mp \frac{1}{2} \sqrt{E_0^2-\alpha^2(1\!\!+\!\!f)^2 + i2\alpha E_0(1\!\!-\!\!f)}
 - i\alpha\frac{1\!\!+\!\!f}{2}.
\label{lambda12}
\end{equation}

\begin{figure}[b]
\includegraphics[width=6.25cm]{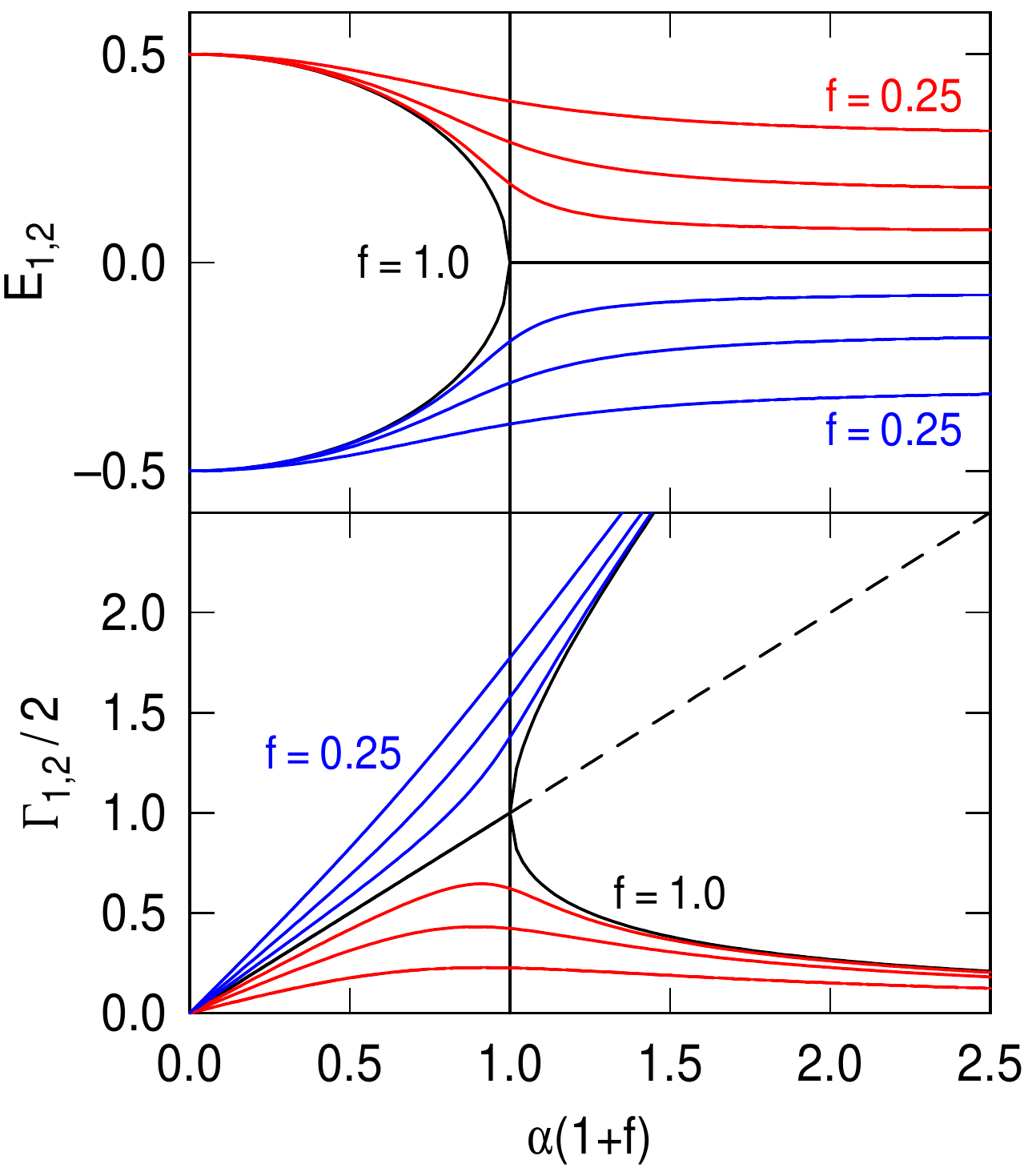}
\caption{
Energy positions $E_{i}= {\rm Re}(\lambda_i)$ and
half widths $\Gamma_{i}/2= -{\rm Im}(\lambda_i)$
of the two-level system (\ref{Heff2}, \ref{lambda12}) as function of the total coupling strength
$\alpha(1\!+\!f)$ for $E_0 = 1$ and $f= \frac{1}{4},\frac{1}{2}, \frac{3}{4}, 1$.
The dashed line denotes $\Gamma_{\rm tot}/2 = \alpha(1\!+\!f)$.
In the special case $f=1$ both resonances are centered at $E=0$ for $\alpha(1\!+\!f) \ge 1$.
} \label{fig_ew2f}
\end{figure}

The dependence of this solution on the value of the coupling parameters $\alpha$ and $f$ is
illustrated in Fig.~\ref{fig_ew2f}. One recognizes that the two resonances attract each other with
increasing coupling strengths $\alpha$. Whereas the width of the lower, more strongly coupled
resonance $\Gamma_1$ increases monotonously with $\alpha$, the width $\Gamma_2$ of the more weakly
coupled upper resonance attains a maximum width around $\alpha(1+f) = 1$. For this coupling, the
combined half width of both resonances $\Gamma_{\rm tot}/2 = \alpha(1+f)$ equals the initial level
spacing $E_0$. In the limit of large coupling, the lower resonance attains the total width
$\alpha(1+f)$ whereas the other one becomes infinitely sharp (see appendix). This narrowing is
known as resonance trapping \cite{Okolow03}.

The construction of the effective Hamiltonian (\ref{Heff}) for the series of
image-potential resonances requires the energy levels of the decoupled series
and the coupling strengths to the metallic continuum.
 These values are obtained from the multiple scattering theory
\cite{Echeni78jp}.
 Within this approach a surface state is viewed as a wave trapped between the
bulk crystal and the surface barrier, similar to the modes of a Fabry-Pérot
interferometer. The repeated scattering at the crystal boundary and the surface
barrier results in an amplitude
\begin{equation}
\frac{1}{1 - r_{\rm B} r_{\rm C} \exp i(\phi_{\rm B} + \phi_{\rm C})}
\label{Psiscatter}
\end{equation}
where $\phi_{\rm B}$ and $\phi_{\rm C}$ denote the phase changes
between incident and reflected waves at the surface barrier and at
the crystal, respectively. $r_{\rm B}$ and $r_{\rm C}$ are the
corresponding reflectivities. In the usual image-state problem
$r_{\rm B} = r_{\rm C} = 1$ and the condition for bound states is
\begin{equation}
\phi_{\rm B} + \phi_{\rm C} = 2\pi n, \quad n=0,1,2,\dots
\label{2pin}
\end{equation}
 In the case of resonances, the loss of flux due to elastic electron
transfer to bulk states results in $r_{\rm C} < 1$. This can be accounted for
in terms of a complex phase $\phi_{\rm C} = \phi_{\rm C}^{'} + i \phi_{\rm
C}^{''}$ with an imaginary component $\phi_{\rm C}^{''} = - \ln r_{\rm C}$
\cite{Echeni78jp}.

Since this scattering model does not take a possible perturbation of the continuum by the
resonances into account, a direct evaluation of eq.~(\ref{Psiscatter}) will not yield the proper
energies and widths of the resonances when $r_{\rm C}$ is small, i.e. in the case of strong
coupling.
 In the weak coupling limit ($r_{\rm C} \rightarrow 1$), however, it can safely be employed to determine
energies and {\em relative} coupling strengths, the only input
needed to construct $\mathcal{H}_{\rm eff}$.
 In linear approximation, the condition (\ref{2pin}) is fulfilled for
\begin{equation}
\frac{\Gamma}{2} \frac{\partial}{\partial E} (\phi_{\rm B} + \phi_{\rm C}^{'}) = \phi_{\rm C}^{''}
\label{GammaE}
\end{equation}
with $\Gamma/2$ being the energy dependent decay rate of the
amplitude (\ref{Psiscatter}).

For the energy dependence of the phase change $\phi_{\rm B}$ at the
image potential we use the WKB approximation \cite{McRae81ss},
$\phi_{\rm B} \simeq [(-8E)^{-1/2} - 1] \pi$.
Above the band gap $\phi^{'}_{\rm C} = \pi$  (see below), and we
obtain a Rydberg series of resonances
\begin{equation}
E_n^0 = - \frac{1}{32 n^2},\ n=1,2,\dots
\label{Rydberg}
\end{equation}
with a decay rate or full width at half maximum given by
\begin{equation}
\Gamma_n^0 = \frac{-\ln r_{\rm C}(E_n^0)}{2\pi} \frac{1}{8 n^3} ,\ n=1,2,\dots
\label{Gamman}
\end{equation}
In the case of a simple metal it is sufficient to restrict oneself to one continuum ($K=1$) like
in the above example of the two-level system. Since Eq.~(\ref{Gamman}) must correspond to the
solution of (\ref{Heff}) in the limit of small coupling, the matrix elements $V_n$ are determined
by the condition $|V_n|^2 = \Gamma_n^0/2$.

 In addition to a $n^3$-dependence, the $\Gamma_n^0$
depend weakly on the quantum number $n$ via the energy dependence of
the reflectivity $r_C$.
 In order to facilitate a general discussion, we will neglect this
weak dependence in the following. This approximation allows us to
introduce the dimensionless parameter
\begin{equation}
\alpha = \frac{1}{\pi} (-\ln r_{\rm C})
\label{alpha}
\end{equation}
as a measure of the overall resonant coupling strength of the whole
series and we obtain
\begin{equation}
|V_n|^2 = \frac{\Gamma_n^0}{2} = \frac{\alpha}{32 n^3}.
\end{equation}
 Please note that in terms of the numerical matrix diagonalization necessary to find
the complex eigenvalues of eq.~(\ref{Heff}), the approximation of a constant
$r_C$, has no advantage and can easily be dropped in calculations for specific
materials. However, it will become apparent below that for the most interesting
cases of strong coupling the energy dependence of $r_{\rm C}$ is indeed
negligible over the limited range of energies $E_n^0$.

\begin{figure}[t]
\includegraphics[width=8.5cm]{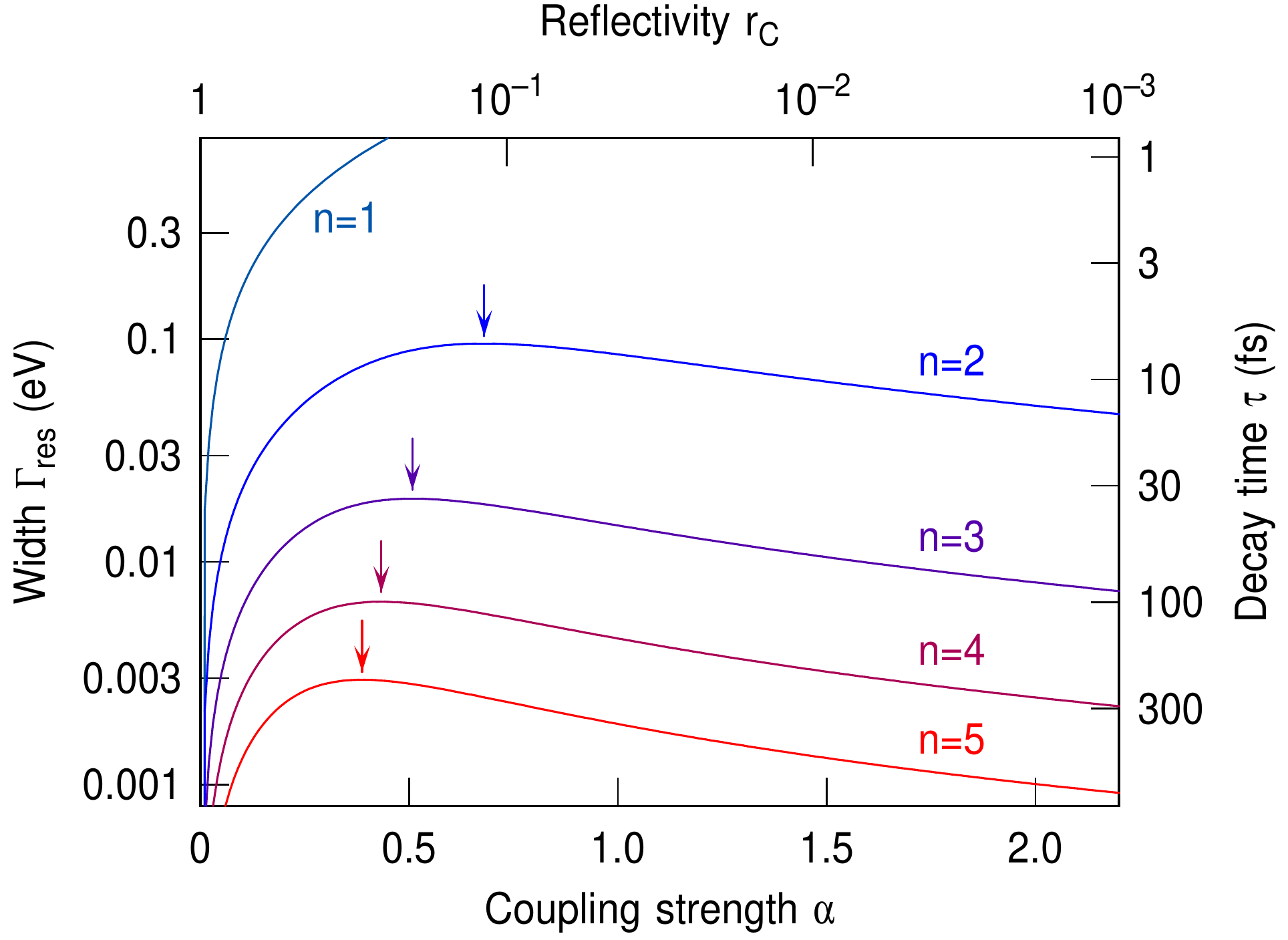}
\caption{
Half widths and decay times of the first five resonances as function of coupling strength
and reflectivity. Arrows indicate coupling strengths where maximum widths are reached.
} \label{fig_iprw}
\end{figure}

\begin{figure}[t]
\includegraphics[width=8.0cm]{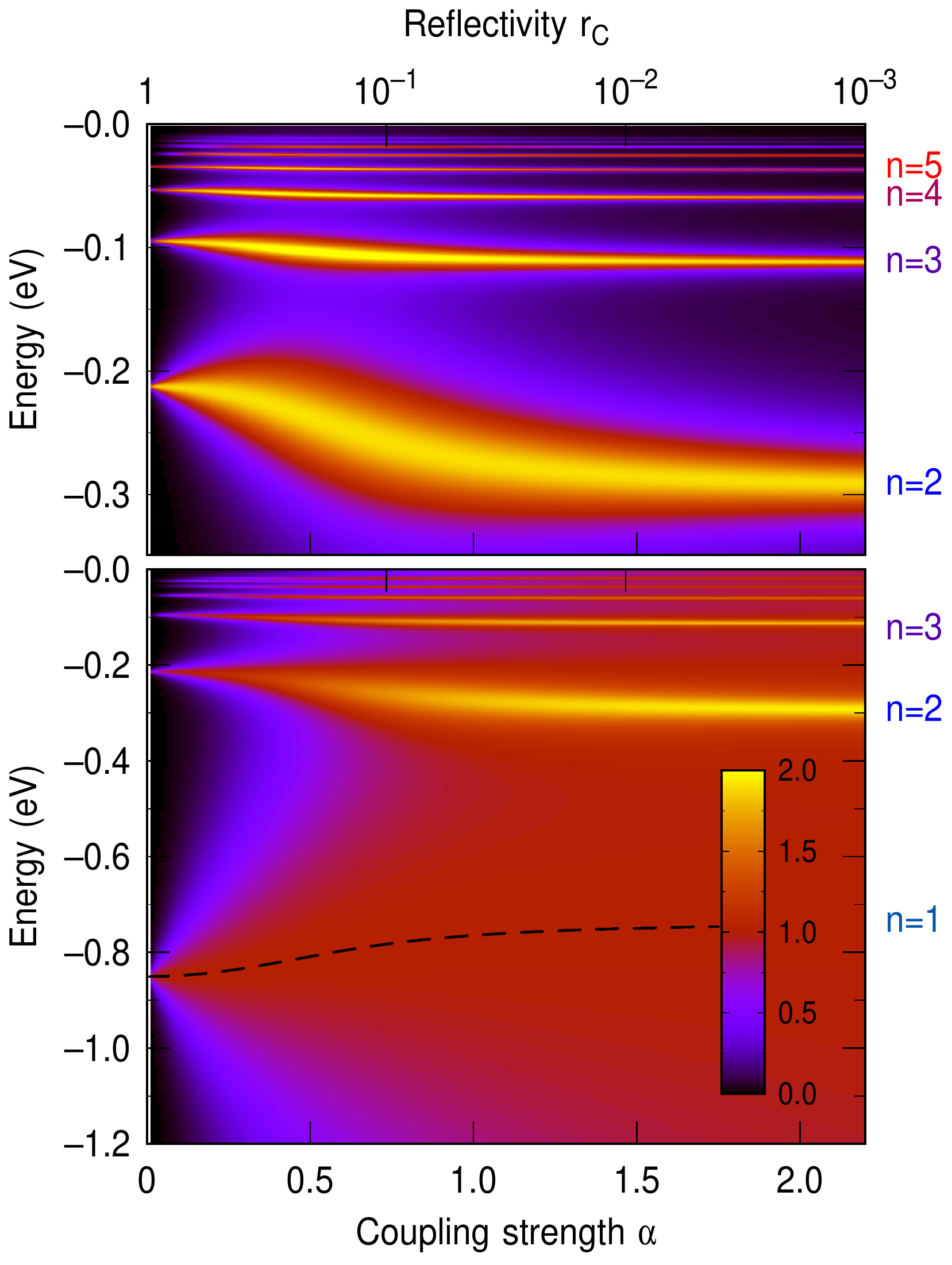}
\caption{
(a) False color contour plot of resonance spectra $n=1,..,5$ as
function of coupling strength/reflectivity. For better visibility
the Lorentzians of the individual resonances are superimposed with
equal amplitudes (not areas); the dashed line indicates the maximum
of $n=1$ resonance.
(b) Resonance spectra $n=2,..,5$ on an expanded
scale; the transition between red and blue colors occurs
approximately at half the maximum amplitude.
} \label{fig_ipre}
\end{figure}

The resulting energies and widths of the first five resonances are
plotted in Figs. \ref{fig_iprw} and \ref{fig_ipre} as function of
the coupling parameter $\alpha$. Specific numerical values of
interest are collected in Table~\ref{tab_ipr}.
 In the limiting case $r_{\rm C} \rightarrow 1$ or $\alpha
\rightarrow 0$ the eigenvalues of (\ref{Heff}) evidently give exactly the
resonance spectrum (\ref{Rydberg}) and (\ref{Gamman}). As expected from the
above analysis of the two-level system, dramatic changes occur as soon as
$\Gamma_n^0/2 + \Gamma_{n+1}^0/2 \simeq \Gamma_n^0 = \alpha/16n^3$ starts to
approach the level spacing $\Delta_n^0 = E_{n+1}^0 - E_n^0 \simeq 1/16n^3$,
i.e. when $\alpha$ approaches unity \cite{fn1}.

\begin{table}[b]
\caption{Limiting values of energies $E_n^0$, $E_n^\infty \equiv
E_n(\alpha\!\rightarrow\!\infty)$, maximum widths $\Gamma_n^{\sf
max}$, minimum lifetimes $\tau_n^{\sf min}$ and corresponding
coupling parameter $\alpha(\Gamma_n^{\sf max})$ for the first five
resonances.}
\begin{ruledtabular}
\begin{tabular}{cccccc}
 & $E_n^0$~(eV) & $E_n^\infty$~(eV) & $\Gamma_n^{\sf max}$~(eV) & $\tau_n^{\sf min}$~(fs) &  $\alpha(\Gamma_n^{\sf max})$  \\
\hline
$n$=1 & -0.8504 & -0.7363 & -       & -       & -       \\
$n$=2 & -0.2126 & -0.2967 & 0.09545 & ~~6.90  & 0.6795  \\
$n$=3 & -0.0945 & -0.1221 & 0.01920 & ~34.28 & 0.5076  \\
$n$=4 & -0.0531 & -0.0593 & 0.00662 & ~99.43 & 0.4329  \\
$n$=5 & -0.0340 & -0.0368 & 0.00295 & 223.10 & 0.3871  \\
\end{tabular}
\end{ruledtabular}
\label{tab_ipr}
\end{table}

In the regime of small coupling ($\alpha \lesssim 0.2$) all
resonances develop uniformly and simply broaden linearly as a
function of $\alpha$ (Fig.~\ref{fig_ipre}). With increasing coupling
the behavior of the $n=1$ resonance and the other resonances become
qualitatively different.
 Whereas the width of $n=1$ increases monotonically and
spreads over the whole spectrum for $\alpha \gtrsim 0.5$, all other resonances
reach a maximum width $\Gamma_n^{\rm max}$ and get narrower again for further
increasing $\alpha$. The higher the quantum number $n$, the sooner
$\Gamma_n^{\rm max}$ is reached (Fig.~\ref{fig_iprw}).
 The positions of the resonances also change as a function of coupling
strength until they reach a limiting value $E_n^{\infty}$ for large
$\alpha$.
 As illustrated by Fig.~\ref{fig_ipre}, the first resonance and the series of all
other resonances effectively attract each other with increasing
coupling. This attraction is larger, the lower the quantum number
$n$. As a result, the resonances $n=2,3,\dots$ spread as a function
of $\alpha$.
 For the maximum coupling strength $\alpha = 2 $ plotted in Fig.~\ref{fig_ipre}, the
energies $E_n$ have reached their limiting values already within 2\%
whereas the deviation from the uncoupled situation is more than 30\%
for $n=2$.

Most importantly, if one compares the widths and the energy separation of the
resonances $n=2,3,\dots$ (Fig.~\ref{fig_ipre}), they are found to be well
separated throughout the range of coupling strengths $\alpha = 0
\rightarrow\infty$. Even around $\alpha = 0.5$ where they reach their maximum
widths $\Gamma_n^{\rm max}$, the combined half-width of neighboring resonances
stays well below their energy difference. {\em This means - with the notable
exception of the first resonance - that the integrity of the Rydberg series is
not destroyed by strong coupling to a continuum}. In contrast, stronger
coupling even leads to a sharper spectrum of the higher resonances as compared
to intermediate coupling.

This, at first glance, surprising result is understood in a straightforward manner by considering
the effective Hamiltonian (\ref{Heff}) in the limit $\alpha \rightarrow \infty$.
 In that case the matrix $VV^{\dagger}$ determines the behavior of
the system and one expects to obtain $K$ states that take almost the
whole coupling and $N-K$ states that are almost decoupled.
 For our case of one continuum ($K$=1), the most strongly coupled
resonance ($n$=1) becomes the fast (open) channel with $\Gamma_1/2
\sim -{\rm Im}\{Tr(\mathcal{H}_{\rm eff})\}$ whereas $n=2,3,\dots$
become long lived (trapped) states.
 A reorganization of the system due to interference of the different
decay channel takes place already for intermediate values of $\alpha$.
 As a consequence, the resonance spectrum never gets totally smeared out as one might intuitively
expect.

In terms of the basic physical processes, the trapping is nothing else than a back feeding of
intensity from the distorted continuum into the more weakly coupled states. This is easily
recognized in the example of the two level system discussed in the appendix. If the system in
state $|a\>$ is coupled to the continuum at $t=0$, the amplitude $a(t)$ will start to decay. At
the same time, the amplitude $b(t)$, which was initially equal to zero, will start to increase in
reaction to the distortion of the continuum (eq. A1). Since the amplitude $b(t)$ has the opposite
phase as $a(t)$, the coupling of state $|b\>$ to the continuum will in turn accelerate the decay
of state $|a\>$.

%

\subsection*{3. Results of Ag(111), Al(100) and Al(111)}

 In our description, the only material-dependent parameter is the
reflectivity $r_{\rm C}$ of a free electron wave at the surface. For surfaces
of simple metals, like Ag(100), Al(100) and Al(111), it can simply be obtained
from the two-band model of the bulk electronic structure \cite{Schiller09prb}.
 For a distance $a$ of lattice planes along the surface normal and an
$sp$-gap at the $\bar\Gamma$-point of $2 V_g$, the potential of the
two-band model is
\begin{equation}
V(z) = -V_0 + V_g e^{igz} + V_g e^{-igz}
\end{equation}
with $g= 2\pi/a$, and $V_0$ chosen such that the vacuum level is at
zero energy. We restrict ourselves to a situation where the
image-potential states are resonant with the upper band
\begin{equation}
E(k) = -V_0 + \big(\frac{g}{2}\big)^2 + k^2 + \sqrt{V_g^2 + g^2k^2}
\end{equation}
The wave functions that solve the Schrödinger equation are Bloch
waves in the crystal ($z<0$) and plane waves in the vacuum ($z>0$)
\begin{subequations}
\begin{eqnarray}
\Psi_{\rm C} &=& \alpha e^{i(\frac{g}{2}-k)z} + \beta e^{-i(\frac{g}{2}+k)z}, \quad z \le 0 \\
\Psi_{\rm vac} &=& A e^{iqz} + B e^{-iqz} , \quad z \ge 0
\end{eqnarray}
\end{subequations} with wave vectors
\begin{subequations}
\begin{eqnarray*}
k &=& \sqrt{{E + V_0} + \big(\frac{g}{2}\big)^2 - \sqrt{(E+V_0)g^2 + V_g^2}} \\
q &=& \sqrt{E+V_0}.
\end{eqnarray*}
\end{subequations}

The reflectivity $r_C$ is the ratio $A/B$ of the in and outgoing
plane wave $\Psi_{\rm vac}$ in the vacuum, obtained from matching
conditions at the interface
 ($\Psi_{\rm C}^{'}/\Psi_{\rm C} |_{z=0} = \Psi_{\rm
vac}^{'}/\Psi_{\rm vac} |_{z=0} $) \cite{fnS1}:
\begin{equation}
r_{\rm C} = \frac{A}{B} =
 \frac{V_g(q-k+\frac{g}{2}) + V'(q-k-\frac{g}{2})}
      {V_g(q+k-\frac{g}{2}) + V' (q+k+\frac{g}{2})} \\
\end{equation}
with
\begin{equation*}
V' = V_g\frac{\beta}{\alpha} = g k + \sqrt{V_g^2 + g^2k^2}.
\end{equation*}

\begin{figure}[b]
\includegraphics[width=7.0cm]{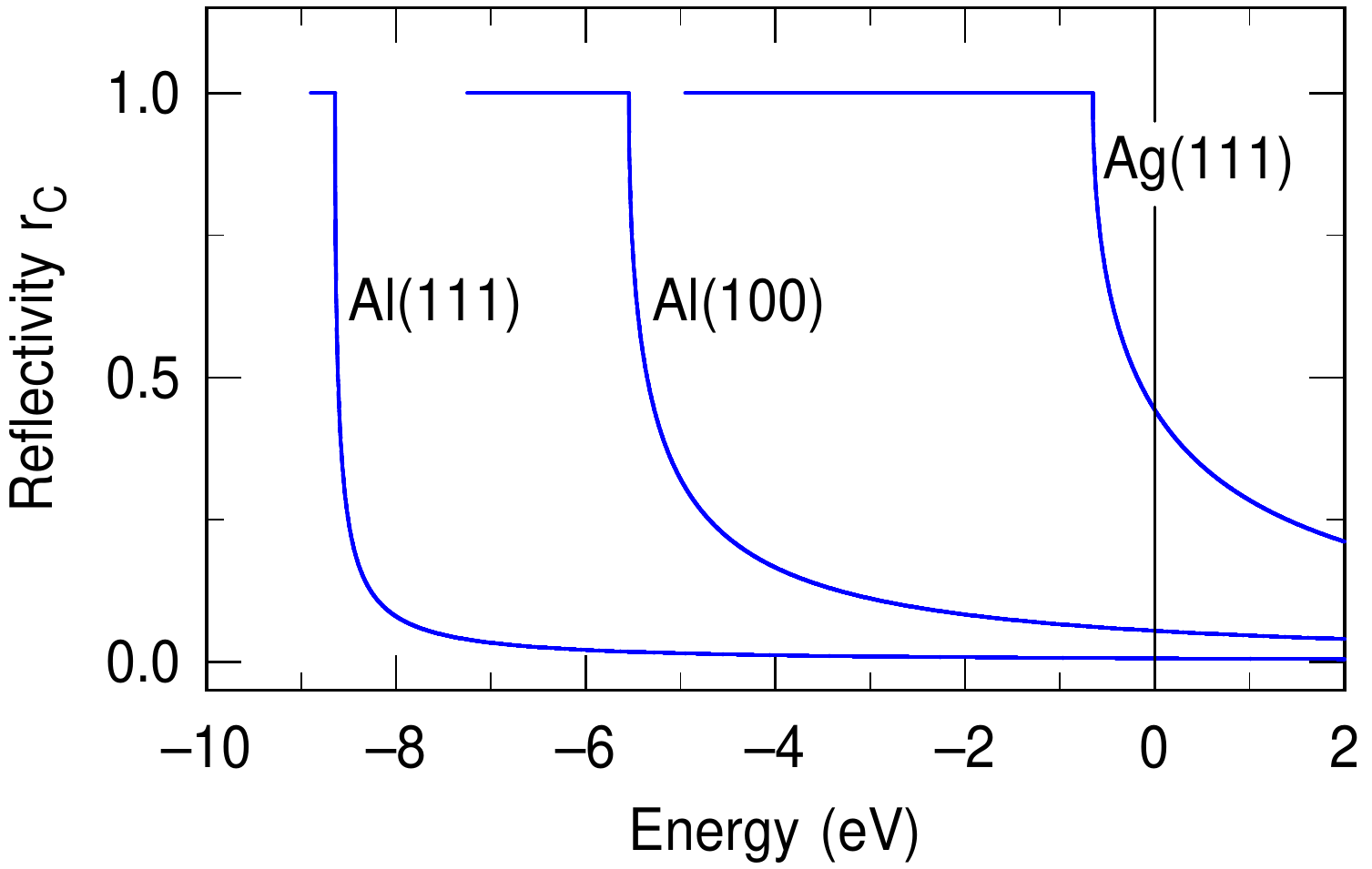}
\caption{
Energy dependence of the reflectivity $r_C$ for Al(111), Al(100), and Ag(111) as calculated from the two-band model.
} \label{fig_rc}
\end{figure}

\begin{figure*}[t]
\begin{center}
\includegraphics[width=7.5cm]{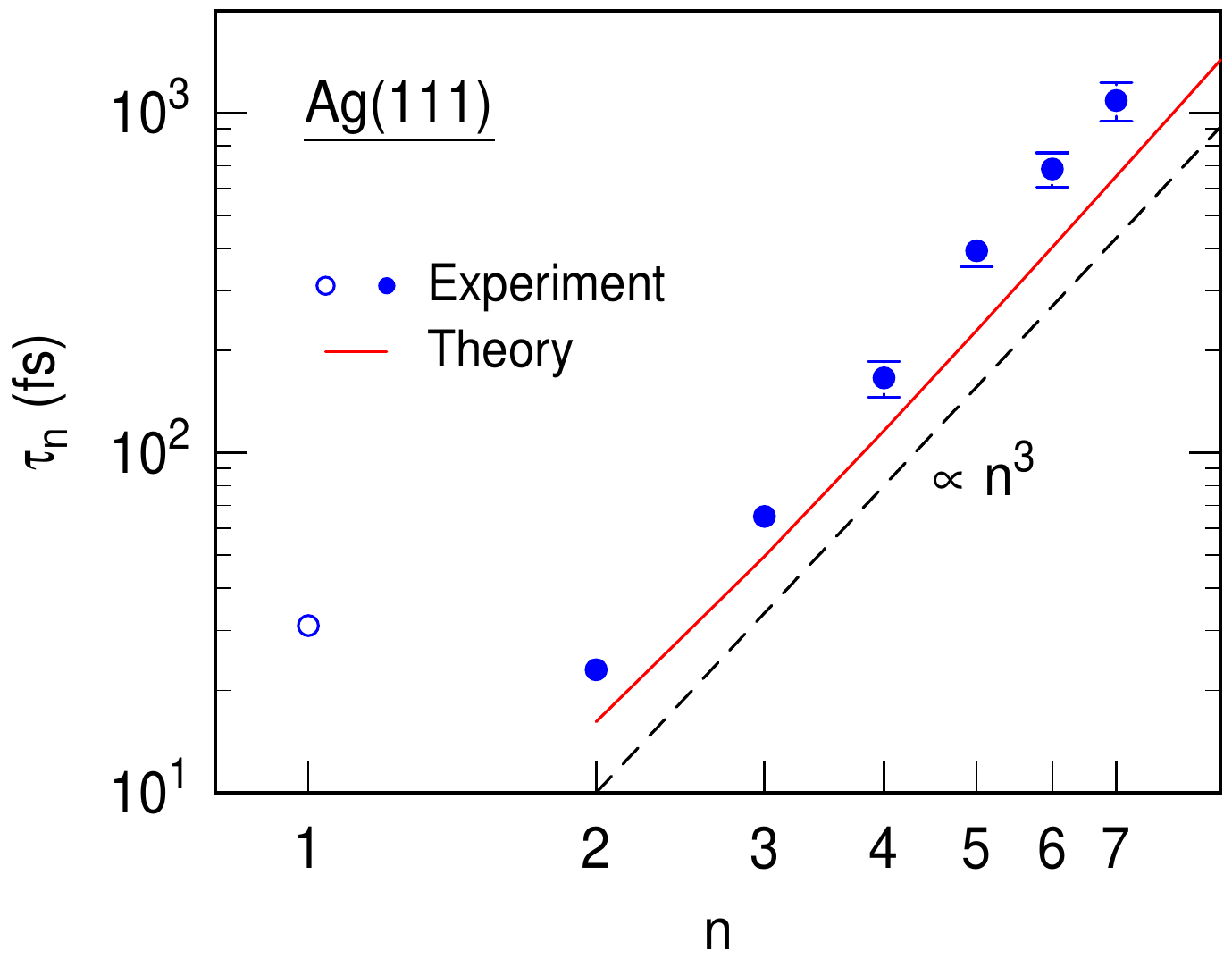} \hspace*{0.25cm} \includegraphics[width=7.5cm]{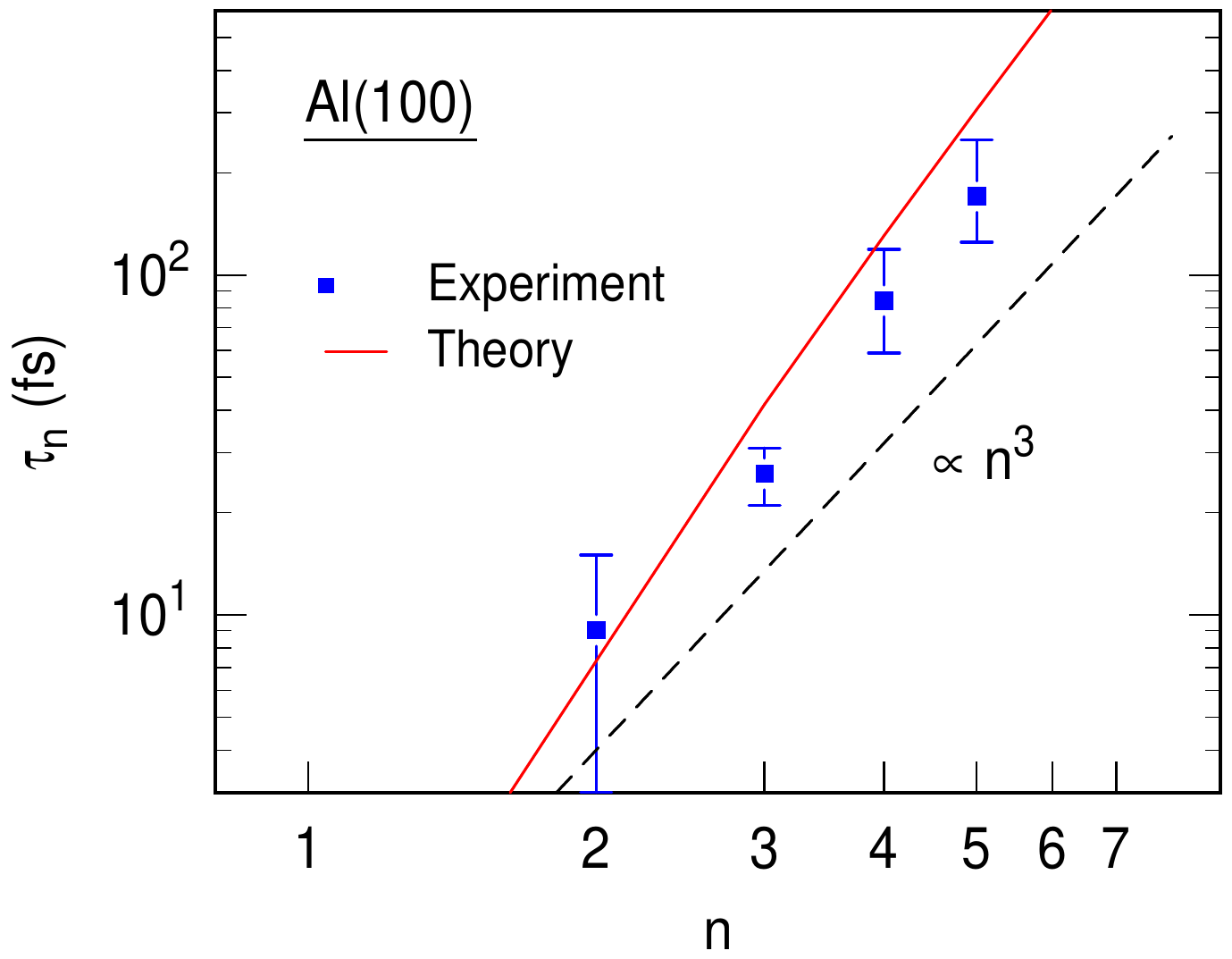}
\caption{
 Comparison of measured lifetimes (full circles) with calculated
values (solid line) using reflectivities $r_C$ from the two-band
model (Table \protect{\ref{tab_rc}}) for Ag(111) (a) and Al(100)
(b); a common coupling parameter of $\alpha=0.92$ was used for
Al(100), whereas the energy dependence of $r_C$ was taken into
account for Ag(111). For this surface, $n=1$ (open circle) is an
inelastically decaying gap state \cite{Marks11prb2}.}
\label{fig_agal}
\end{center}
\end{figure*}

 Parameters used to describe the three surfaces Ag(111), Al(100), and
Al(111) and the resulting reflectivities $r_C$  and coupling parameters $\alpha
= - \pi^{-1}\ln r_C $ for $E=E_{\rm vac} \equiv 0$ are collected in
Table~\ref{tab_rc}.
 Explicit values of the calculated resonance energies and lifetimes are given in
Table \ref{tabew}.
 $r_{\rm C}$ is equal to one at the edge of the upper band
($k=0$) and decreases monotonously as the energy increases. This decrease is more rapid, the
smaller the value of $V_g$ (Fig.~\ref{fig_rc}).

\begin{table}[b]
\caption{Reflectivities at the Vacuum Energy $r_{\rm C}(E\!=0\!)$
and coupling strengths $\alpha$ for Ag(111), Al(100) and Al(111);
parameters have been derived from the values given in
Ref.~\cite{Chulkov99ss2} (band gap: $E_{\rm gap} = 2V_g$, upper band
edge: $E_{\rm up} = E(k=0)$).}
\begin{ruledtabular}
\begin{tabular}{llrrrll}
Surface & $a$~(\AA) & $V_g$~(eV) & $V_0$~(eV) & $E_{\rm up}$~(eV) & $r_{\rm C}$ & $\alpha$ \\
\hline
Ag(111) & $4.085/\sqrt{3}$ & 2.150  &  9.3619 & -0.4527 & 0.51   & 0.22  \\
Al(100) & $4.049/\sqrt{2}$ & 0.840  & 10.9723 & -5.5450 & 0.054  & 0.92  \\
Al(111) & $4.049/\sqrt{3}$ & 0.125  & 15.6485 & -8.6425 & 0.0059 & 1.63  \\
\end{tabular}
\end{ruledtabular}
\label{tab_rc}
\end{table}

Experimental data for the series of image-potential resonances are
available for Ag(111) \cite{Marks11prb2} and Al(100)
\cite{Winter11prl}. Fig.~\ref{fig_agal} shows that the present
theory is able to describe the measured lifetimes within a factor of
two.
 Apparently, the two-band model predicts somewhat too small values of $r_C$.
In the case of Ag(111), this results in shorter lifetimes than the
experimental ones. In the case of Al(100), on the contrary, too small
values of $r_C$ overestimate the lifetimes of the high-$n$ states
that are already effectively trapped.

 For Ag(111) $n=1$ is still a gap state, $r_C$ is 50\% near the vacuum energy.
The corresponding coupling parameter is $\alpha = 0.22$. Trapping
will thus be effective only for resonances with higher quantum
numbers.
 For Al(100) the reflectivity is only 5\% and the coupling
parameter is $\alpha = 0.92$.
 The theory predicts a very broad first resonance and widths of the
high-$n$ resonances that are substantially reduced as compared to
the model of independent decay (Fig.~\ref{fig_iprw}). The
quantitative comparison with the experimental data of
Refs.~\cite{Marks11prb2} and \cite{Winter11prl} shows good
quantitative agreement for both surfaces, particularly if one
considers that we neglect other decay channels and make use of the
most simple estimate for $r_C$.
 Al(111) is a surface very close to the idealized jellium model with
a very low electron reflectivity ($r_{\rm C} = 0.6\%$). With a
resulting coupling parameter $\alpha = 1.63$ the present theory
predicts high-$n$ resonances that are sharper and more long-lived
than those of Al(100) and even those of Ag(111). Electrons in these
resonances are effectively trapped (Fig.~\ref{fig_iprw}).

\begin{table}[b]
\caption{Energies $E_n$ and decay times $\tau_n$ of the first seven
image-potential resonances of Ag(111), Al(100) and Al(111).}
\begin{ruledtabular}
\begin{tabular}{c|rrr|rrr|rr}
 & & \!\!Ag(111) && & \!\!Al(100) && & \!\!Al(111) \\
 & $E_n$~(eV) & $\tau_n$~(fs) && $E_n$~(eV) & $\tau_n$~(fs) && $E_n$~(eV) & $\tau_n$~(fs) \\
\hline
$n$=1 &    -   &   -   && -0.767 &    0.4 && -0.745 &    0.2 \\
$n$=2 & -0.211 &  16.2 && -0.270 &    7.3 && -0.288 &   10.8 \\
$n$=3 & -0.095 &  49.6 && -0.109 &   41.6 && -0.111 &   65.9 \\
$n$=4 & -0.054 & 116.2 && -0.058 &  130.7 && -0.059 &  221.4 \\
$n$=5 & -0.034 & 229.2 && -0.037 &  307.3 && -0.037 &  502.9 \\
$n$=6 & -0.024 & 402.6 && -0.025 &  608.6 && -0.025 & 1003.6 \\
$n$=7 & -0.018 & 651.0 && -0.018 & 1075.2 && -0.018 & 1782.9 \\
%
\end{tabular}
\end{ruledtabular}
\label{tabew}
\end{table}

 It must be pointed out that the result for Al(111) stands in marked
contradiction with some of the established knowledge from the
literature. Previous experimental and theoretical work agree on the
presence of just one structureless feature near the vacuum level for
Al(111) arising from image-potential resonances
\cite{Heskett87prb,Yang93prl,Bulovic94jvsta,Lindgren89prb,Papadia90prb,Fratesi03prb}.
 However, the experimental resolution of Refs.~\cite{Heskett87prb,Yang93prl,Bulovic94jvsta}
was less than 200 meV and not sufficient to resolve high-$n$ resonances.
 A reinvestigation of well-defined Al(111) with improved experimental possibilities would therefore
be highly desirable.
 The Al(111) surface would also provide interesting opportunities to decouple the
resonances by adlayers. The present theory clearly predicts that moderate
decoupling should lead to shorter lifetimes of the high-$n$ resonances, and not
to longer ones as for image-potential states at gaps \cite{Gudde06cr}.

\subsection*{5. Conclusions}

 In conclusion, we have shown that electrons in energetically adjacent states,
strongly coupled to a continuum at a metal interface, do not
delocalize independently from each other.
 Interference effects can lead to extremely long lifetimes of some states while
others experience an accelerated decay.
 Criterium for strong coupling is a significant overlap of the broadened levels,
not the absolute coupling strength. One expects it to be fulfilled
in a variety of situations in nanoscience, catalysis or molecular
electronics, e.g. when small clusters or organic molecules with many
close lying levels interact with a substrate.
 As for the Rydberg series of image-potential resonances, our theory strengthens the
prediction of Echenique and Pendry \cite{Echeni78jp}, who stated that the whole series of states
would be observable whenever the first two ones are separated.
 With the notable exception of the first resonance, a series of
clearly separated resonances should exist for all coupling strengths, i.e. on
any well-defined metal surface.

\subsection*{Acknowledgement}

We thank J.-P.~Gauyacq, T.F.~Heinz, P.~Thomas and M.~Winter for
valuable discussions and acknowledge funding by the Deutsche
Forschungsgemeinschaft through SFB 1083.

\appendix
\section{Effective Hamiltonian of the Two-Level System}

 In the literature, the effective Hamilton operator $\mathcal{H}_{\rm eff}$ for the
general case of $n$ levels interacting with several continua is
usually obtained by means of the Feshbach projection operator
formalism \cite{Feshbach58,Desout95jp,Rotter09jp}.
 Much of the essential physics, however, is already revealed by considering two levels
coupled to one structureless continuum \cite{Devdar76jetp}. In the following we
give a simple derivation of $\mathcal{H}_{\rm eff}$ for the two-level system
(\ref{Heff2}) and evaluate its properties. In this way, we hope to make the
approach more easily accessible to the general surface science reader.


 The Hamilton operator of two electronic states $|a\>$ and $|b\>$ with energies
$E_a$ and $E_b$ coupled to a continuum by an interaction $V$ is
\begin{equation*}
H = H_0 + V = E_a|a\>\<a| + E_b|b\>\<b| + \sum_\mu \omega_\mu |\mu\>\<\mu| + V.
\end{equation*}
We consider here only an interaction between the states $|a\>$ and
$|b\>$ and the continuum states $|\mu\>$
\begin{equation*}
\<a|V|\mu\> = V_{a},\quad \<b|V|\mu\> = V_{b}.
\end{equation*}
 This interaction is assumed to be energy independent; all other matrix
elements $\<|V|\>$ vanish.
 The time dependence of the wave function of the full system
\begin{equation*}
|\Psi(t)\> = a(t)|a\> + b(t)|b\> + \sum_\mu \hat{c}_\mu(t)e^{-i\omega_\mu t} |\mu\>
\end{equation*}
 is determined by the Schrödinger equation $H|\Psi\> = i\partial_t |\Psi\>$
 which reads in matrix form
\begin{equation*}
i\partial_t
\begin{pmatrix}
a \\
b \\
\hat{c}_1 e^{-i\omega_1 t} \\
\vdots \\
\hat{c}_N e^{-i\omega_N t}
\end{pmatrix}
=
\begin{pmatrix}
    E_a &      & V_a      & \cdots & V_a \\
       &  E_b   & V_b     & \cdots &  V_b \\
    V_a &  V_b   & \omega_1 &      \\
 \vdots & \vdots &          & \ddots \\
    V_a &  V_b   &          &        & \omega_N
\end{pmatrix}
\begin{pmatrix}
a \\
b \\
\hat{c}_1 e^{-i\omega_1 t} \\
\vdots \\
\hat{c}_N e^{-i\omega_N t}
\end{pmatrix}
\end{equation*}
 or as a set of coupled differential equations
\begin{subequations}
\begin{eqnarray}
i\partial_t a(t) &=& E_a a(t) + V_a \sum_\mu \hat{c}_\mu(t) e^{-i\omega_\mu t}  \label{PDG1-3a} \\
i\partial_t b(t) &=& E_b b(t) + V_b \sum_\mu \hat{c}_\mu(t) e^{-i\omega_\mu t} \label{PDG1-3b} \\
i\partial_t \left[ \hat{c}_\mu(t) e^{-i\omega_\mu t}\right]  &=&
V_a a(t) + V_b b(t) + \omega_\mu\hat{c}_\mu(t) e^{-i\omega_\mu t}. \nonumber \\[-5pt]
&&
\qquad\qquad
\label{PDG1-3c}
\end{eqnarray}
\end{subequations}

 We are only interested in the time dependence of the states $|a\>$ and $|b\>$ due
to their interaction with the continuum, not in the evolution of the
continuum states themselves. We thus eliminate $\hat{c}_\mu$ from
(\ref{PDG1-3a}) and (\ref{PDG1-3b}) by integrating (\ref{PDG1-3c})
and obtain
\begin{equation*}
 \hat{c}_\mu(t)  = -i\int dt' [V_a a(t') + V_b b(t')] e^{i\omega_\mu t'}.
\end{equation*}
 We further assume the continuum states $|\mu\>$ to be equally spaced
and to extend from $-\infty$ to $+\infty$. Then the sums that appear in (\ref{PDG1-3a}) and
(\ref{PDG1-3b}) simply become
\begin{eqnarray*}
\sum_\mu \hat{c}_\mu(t) e^{-i\omega_\mu t} &=& \int d\omega\, \hat{c}(\omega,t) e^{-i\omega t} \\[-5pt]
&=& -i \int d\omega \int dt' [V_a a(t') + V_b b(t')] e^{i\omega (t'-t)} \\[5pt]
&=& -i 2\pi [V_a a(t) + V_b b(t)]
\end{eqnarray*}
and we arrive at two coupled differential equations for the amplitudes $a(t)$ and $b(t)$
\begin{equation}
i\partial_t
\begin{pmatrix}
a(t) \\[5pt]
b(t)
\end{pmatrix}
=
\begin{pmatrix}
    E_a - i 2\pi V_a^2   &     - i 2\pi V_a V_b     \\[5pt]
        - i 2\pi V_a V_b & E_b - i 2\pi V_b^2
\end{pmatrix}
\begin{pmatrix}
a(t) \\[5pt]
b(t)
\end{pmatrix}.
\label{abeff}
\end{equation}
We identify the matrix in (\ref{abeff}) with the effective Hamiltonian for the
two-level system (\ref{Heff2}) with $E_{a,b} = \mp E_0/2$, $2\pi V_a^2 =
\alpha$, and $2\pi V_b^2 = \alpha f$.

 One recognizes that the coupling to the continuum not only introduces a finite
width of the levels, given by $2\pi V_a^2=\alpha$ and $2\pi
V_b^2=\alpha f$. In addition the two levels appear to interact with
one another as described by the imaginary off-diagonal terms $i 2\pi
V_a V_b = i \alpha \sqrt f$.
 The derivation clearly reveals that this interaction is not a
direct one, but one mediated by the continuum. The interaction with the levels $|a\>$ and $|b\>$
disturbs the continuum states $|\mu\>$. This time-dependent disturbance in turn acts back on
$|a\>$ and $|b\>$. This interaction becomes stronger with a stronger coupling to the continuum,
and it is directly linked to the level widths.
 The derivation also shows that the effective Hamiltonian (\ref{Heff2}) will be a good
approximation for the full Hamiltonian as far as the levels are concerned, when the continuum is
structureless and extends well beyond the energy spread of the broadened levels. This requirement
is fulfilled in the case of the image-potential resonances considered in this work.


The effective Hamiltonian (\ref{Heff2}) of the two-level system with
energies $\mp E_0/2$, overall coupling parameter $\alpha$ and
relative coupling strength $f$ ($ 0 < f \le 1$) has two complex
eigenvalues
\begin{equation}
 \lambda_{1,2} =
 \mp \frac{1}{2} \sqrt{E_0^2-\alpha^2(1\!\!+\!\!f)^2 + i2\alpha E_0(1\!\!-\!\!f)}
 - i\alpha\frac{1\!\!+\!\!f}{2}. \tag{\ref{lambda12}, repeated}
\end{equation}

 The resulting resonance energies and widths can be written in the form
\begin{subequations}
\begin{eqnarray}
E_{1,2} &=& {\rm Re}\, \lambda_{1,2} = \mp \frac{1}{2} \rho \cos\frac{\varphi}{2} \\
\Gamma_{1,2} &=& -2\, {\rm Im}\, \lambda_{1,2} = \alpha(1+f) \pm \rho \sin\frac{\varphi}{2}
\end{eqnarray}
\end{subequations}
 with $\rho = \rho(E_0,f,\alpha)$ and
 $\varphi = \varphi(E_0,f,\alpha)$.
 Obviously the total energy of the coupled system
$E_{\rm tot} = E_1 + E_2 = 0$ does not change with the coupling
parameter $\alpha$ while the total decay rate simply grows linearly
with $\alpha$
\begin{equation}
\Gamma_{\rm tot} = \Gamma_1+ \Gamma_2= 2\alpha(1+f).
\end{equation}
With increasing values of $\alpha$, the energy separation of the two levels $\Delta = E_2 - E_1$
decreases. If one expands the solution into powers of $\alpha$, one finds that the attraction is
initially quadratic in $\alpha$
\begin{equation*}
\Delta(\alpha\rightarrow 0) = E_0 + \frac{f}{E_0}\alpha^2 - \dots
\end{equation*}
and reaches a limiting value that depends on the relative coupling
strength $f$
\begin{equation*}
 \Delta(\alpha\rightarrow\infty)= E_0\frac{1-f}{1+f} + \mathcal{O}(\frac{1}{\alpha^2}).
\end{equation*}
For the degenerate case $f=1$ the attraction is strictly quadratic
until the two levels coincide for $2\alpha \ge 1$.
 While the system remains completely symmetrical in terms of the
energies of the two resonances, the initial asymmetry of the half
widths is seen to increase as a function of $\alpha$.
\begin{eqnarray*}
\frac{\Gamma_{1}}{2} (\alpha\rightarrow 0) &=& \alpha + \frac{\alpha^3}{E_0^2}f(1-f) + \dots \\
\frac{\Gamma_{2}}{2} (\alpha\rightarrow 0) &=& f \alpha - \frac{\alpha^3}{E_0^2}f(1-f) + \dots
\end{eqnarray*}
The width $\Gamma_{1}$ of the more strongly coupled resonance
increases monotonously. The width $\Gamma_{2}$, however, reaches a
maximum value $\Gamma_{2,\rm max}$ and goes to zero in the limit of
infinite coupling.
\begin{eqnarray*}
\frac{\Gamma_{1}}{2} (\alpha\rightarrow\infty) &=& (1+f)\alpha - \frac{E_0^2}{\alpha}\frac{f}{(1+f)^3} + \dots \\
\frac{\Gamma_{2}}{2} (\alpha\rightarrow\infty) &=& \frac{E_0^2}{\alpha}\frac{f}{(1+f)^3} + \dots
\end{eqnarray*}
\label{Gammalimits2}
 The maximum of $\Gamma_{2}$ is reached approximately when the real
part of the argument under the square root in eq.~(3) becomes zero.
This is the case for
\begin{equation}
\alpha_{\rm max}^0  = \frac{E_0}{1+f}
\label{alphamax}
\end{equation}
i.e.~for a value where the total decay rate $\Gamma_{\rm tot}/2 =
\alpha_{\rm max}^0(1+f)$ is equal to the initial level spacing
$E_0$.
 For this coupling strength we have
\begin{subequations}
\begin{eqnarray}
\Delta(\alpha_{\rm max}^0) &=& E_0\sqrt{\frac{1-f}{1+f}} \\
\label{Deltamax}
\Gamma_{2}(\alpha_{\rm max}^0) &=& E_0 - \Delta(\alpha_{\rm max}^0)
\label{Gammamax}
\end{eqnarray}
\end{subequations}

These results concerning the maximum of $\Gamma_{2}$ are exact for
$f=1$. In this degenerate case, the widths of both levels increase
strictly linear with $\alpha$ as long as $2\alpha \le 1$. For
$2\alpha > 1$, when their energies coincide, one of them increases
faster, while the other one decreases and approaches zero (see also
Fig.~1 in the main text).
 For a ratio $f < 1$ eq.~(\ref{alphamax}) slightly overestimates the
coupling parameter $\alpha$ for which the maximum of $\Gamma_{2}$ is
reached. The error, however, is less than 12\% and less than 4\% in
expression (\ref{Gammamax}) which slightly underestimates
$\Gamma_{2,\rm max}$.

\bigskip


\begin{thebibliography}{10}
\bibitem{fn1} According to our definition,the dimensional parameter $\alpha$ corresponds
  exactly to the ratio $\Gamma_n/\Delta_n$ of the noninterfering system for
  large $n$.
\bibitem{fnS1} The derived reflectivities are independent from a specific postition of the
  bounary $z_0$ in the framework of this model.
\end{thebibliography}


\end{document}